\documentclass[10pt]{article}
\usepackage{amsmath,amssymb,tikz}
\usepackage{graphicx}
 \allowdisplaybreaks
\long\def\symbolfootnote[#1]#2{\begingroup%
\def\thefootnote{\fnsymbol{footnote}}\footnote[#1]{#2}\endgroup}

\def\({\left(} \def\){\right)}
\def\[{\left[} \def\]{\right]}

\def\go{\mathrel{\mathop g^{\scriptscriptstyle{(1)}}}{}\!}
\def\gs{\mathrel{\mathop g^{\scriptscriptstyle{(2)}}}{}\!}

\newcommand{\be}{\begin{equation}}
\newcommand{\ee}{\end{equation}}
\newcommand{\bea}{\begin{eqnarray}}
\newcommand{\eea}{\end{eqnarray}}
\newcommand{\ba}{\begin{eqnarray}}
\newcommand{\ea}{\end{eqnarray}}

\newcommand{\beq}{\begin{equation}}
\newcommand{\eeq}{\end{equation}}
\newcommand{\beqa}{\begin{eqnarray}}
\newcommand{\eeqa}{\end{eqnarray}}
\newcommand{\beqar}{\begin{eqnarray*}}
\newcommand{\eeqar}{\end{eqnarray*}}

\newcommand{\veps}{\varepsilon}

\def\de{\delta}

\newcommand{\aei}{\it Max Planck Institute for Gravitational Physics
(Albert Einstein Institute)\\ Am M\"uhlenberg 1, 14476 Golm,
Germany}

\newcommand{\itp}{\it State Key Laboratory of Theoretical Physics, Institute of Theoretical Physics, \\ Chinese Academy of Sciences, Beijing, 100190}

\newcommand{\kitpc}{\it Kavli Institute for Theoretical Physics China,\\
Chinese Academy of Sciences, Beijing, 100190}
\begin{document}
\thispagestyle{empty}
\begin{center}

~\vspace{20pt}

{\Large\bf A note on the resolution of the entropy discrepancy}

\vspace{25pt}

Yue Huang\symbolfootnote[1]{Email:~\sf huangyue@itp.ac.cn}${}^\dag{}$,   Rong-Xin Miao\symbolfootnote[3]{Email:~\sf rong-xin.miao@aei.mpg.de}

\vspace{10pt}${}^\ast{}$\itp\\
\vspace{10pt}${}^\dag{}$\kitpc\\
\vspace{10pt}${}^\ddag{}$\aei

\vspace{2cm}

\begin{abstract}
It was found by Hung, Myers and Smolkin that there is entropy discrepancy for the CFTs in 6-dimensional space-time, between the field theoretical and the holographic analysis. Recently, two different resolutions to this puzzle have been proposed. One of them suggests to utilize the anomaly-like entropy and the generalized Wald entropy to resolve the HMS puzzle, while the other one initiates to use the entanglement entropy which arises
from total derivative terms in the Weyl anomaly to explain the HMS mismatch. We investigate these two proposals carefully in this note. By studying the CFTs dual to Einstein gravity, we find that the second proposal can not solve the HMS puzzle. Moreover, the Wald entropy formula is not well-defined on horizon with extrinsic curvatures, in the sense that, in general, it gives different results for equivalent actions.  
\end{abstract}

\end{center}

 \newpage
 \setcounter{footnote}{0}
 \setcounter{page}{1}

\tableofcontents

\section{Introduction}

Hung, Myers and Smolkin (HMS) found that the field theoretical and holographic logarithmic terms of entanglement entropy do not match for 6d CFTs \cite{Hung}. For simplicity, we denote this entropy discrepancy as  'HMS puzzle' or 'HMS mismatch' in this note. Recently, two different approaches are proposed to resolve this entropy discrepancy. One of them suggests to utilize the anomaly-like entropy and the generalized Wald entropy derived from the Weyl anomaly to solve the HMS puzzle \cite{Miao1}. While the other one initiates to use the entropy which arises
from total derivative terms in the Weyl anomaly to explain the HMS mismatch \cite{Astaneh1,Astaneh2}.
The question that which proposal is correct is an important problem. We clarify this issue in this note.

It is worth to point out that the results in \cite{Astaneh1,Astaneh2} are crucially based on the regularization given in \cite{Fursaev}. If the Lewkowycz-Maldacena regularization \cite{Maldacena,Dong1} is applied instead, the entropy of covariant total derivatives vanishes\cite{Dong2}. This implies that the proposal of \cite{Astaneh1,Astaneh2} is unreliable. In this note, we give a solid proof that the approach in \cite{Astaneh1,Astaneh2} actually fails in solving the HMS puzzle.

It is counterintuitive that total derivative terms in the Weyl anomaly, arising from cohomologically trivial solutions
to the Wess-Zumino consistency conditions,  contribute to non-zero entropy. Given this fact, the logarithmic term of entanglement entropy of CFTs would depend on the approaches of regularization \cite{Astaneh1,Astaneh2}. However, entropy is physical and thus should be independent of the choices of regularization. The authors of \cite{Astaneh1,Astaneh2} argued that this is not a problem for 4d CFTs,  since no total derivative term appears in the holographic Weyl anomaly in 4d space-time \cite{Bhattacharyya}. Nevertheless, total derivatives do appear in the holographic Weyl anomaly in 6d space-time. The authors of \cite{Astaneh1,Astaneh2} propose to utilize the entropy arising from these total derivative terms to explain the HMS mismatch. They did not take into account all the total derivative terms but only part of them to resolve the HMS mismatch \cite{Astaneh2}.

In this note, we apply the method of \cite{Astaneh1,Astaneh2} to investigate the logarithmic term of entanglement entropy for 6d CFTs dual to Einstein gravity. In contrast to \cite{Astaneh2}, we examine all the total derivative terms in the holographic Weyl anomaly and find that the field theoretical result does not match the holographic analysis. Thus, the proposal of \cite{Astaneh1,Astaneh2} does not resolve the HMS puzzle \cite{Hung}. This is the main new result of this note.

We also find that the Wald entropy formula
\begin{eqnarray}\label{WaldentropyIn}
S_{\text{Wald}}=-2\pi \int_{\Sigma}dx^{D-2}\sqrt{h}\frac{\delta L}{\delta R_{ijkl}} \epsilon_{ij}\epsilon_{kl}
\end{eqnarray}
is not well-defined on the horizon with non-zero extrinsic curvatures. In general, it is inconsistent with the Bianchi identities. It turns out that only the total gravitational entropy, which consists of Wald entropy \cite{Wald}, the anomaly-like entropy \cite{Fursaev,Dong1,Camps} and the generalized Wald entropy \cite{Miao1}, is well-defined. Similar to the Weyl anomaly, the anomaly-like entropy arises from the would-be logarithmic terms in the gravitational action \cite{Dong1}. Notice that it only appears in the higher curvature gravity rather than Einstein gravity. In addition to Wald entropy \cite{Wald} and the anomaly-like entropy \cite{Fursaev,Dong1,Camps}, a new component of entropy appears in general higher derivative gravity. It is named as `generalized Wald entropy' in \cite{Miao1} because of its similarity to Wald entropy. In this note, we mainly focus on the total gravitational entropy and denote it as the total entropy below for simplicity.

The note is organised as follows. In Sect. 2, we briefly review the HMS entropy discrepancy \cite{Hung} and two possible resolutions \cite{Miao1, Astaneh1,Astaneh2}. In Sect. 3, the method in \cite{Astaneh1,Astaneh2} is  employed to calculate the logarithmic term of entanglement entropy for 6d CFTs dual to Einstein gravity, while in Sect. 4 we apply the method of \cite{Miao1}. It turns out that it is the proposal of \cite{Miao1} rather than the one of \cite{Astaneh1,Astaneh2} that can resolve the HMS puzzle. Further evidences for this conclusion are provided in Sect.5. In sect.6, we show that, in general, Wald entropy gives different results for equivalent actions, while the total entropy is indeed well-defined.  We conclude with some discussions in Sect. 7.

\section{The HMS entropy discrepancy}

In this section, we briefly review the HMS entropy discrepancy \cite{Hung}. It was found by Hung, Myers and Smolkin that the logarithmic term of entanglement entropy derived from the field theoretical approach does not agree with the holographic result for 6d CFTs \cite{Hung}. For simplicity, they only focus on the case with zero extrinsic curvature. 

In the field theoretical approach, the logarithmic term of entanglement entropy can be derived by taking the Weyl anomaly as a gravitational action and then calculating the `entropy' of this `action'\cite{Hung, Fursaev}. It turns out that this 'entropy'  equals to the logarithmic term of entanglement entropy for CFTs \cite{Hung, Fursaev}. In 6-dimensional space-time, the Weyl anomaly of CFTs takes the following form
 \be
\langle\,T^i{}_i\,\rangle=\sum_{n=1}^3 B_n\, I_n + 2 A \, E_6 +\nabla_i \hat{J}^i,
\label{trace6}
 \ee
where $B_i, A$ are central charges, $E_6$ is the Euler density, $\nabla_i \hat{J}^i$ are total derivative terms and $I_i$ are conformal invariants given by
 \bea
I_1&=&C_{k i j l}C^{i m n j}C_{m~~\,n}^{~~k l} ~, \qquad
I_2=C_{i j}^{~~k l}C_{k l}^{~~m n}C^{~~~i j}_{m n}~, \\
I_3&=&C_{i k l m}(\nabla^2 \, \de^{i}_{j}+4R^{i}{}_{j}-{6 \over 5}\,R
\,
\de^{i}_{j})C^{j k l m}\,.\label{trace6x}
 \eea
For the entangling surfaces with the rotational symmetry, only Wald entropy contributes to holographic entanglement entropy. Thus, we have \cite{Hung}
 \beq
S = \log(\ell/\delta)\, \int d^{4}y\sqrt{h}\,\left[\,2\pi\,
\sum_{n=1}^3 B_n\,\frac{\partial I_n}{\partial R^{i j}{}_{k
l}}\,\tilde{\veps}^{i j}\,\tilde{\veps}_{k l} +2\,A\,
E_{4}\,\right]_{\Sigma}\,,
 \label{Waldformula8z}
 \eeq
where
 \bea
  \frac{\partial I_1}{\partial R^{i j}{}_{k l}}\,\tilde{\veps}^{i
j}\,\tilde{\veps}_{k l} &=& 3 \left(C^{j m n k}\, C_{m~~n}^{~~i
l}\,\tilde\veps_{i j}\,\tilde\veps_{k l} - \frac{1}{4}\,
C^{iklm}\,C^{j}_{~kl m }\,\tilde{g}^{\perp}_{i j} + \frac{1}{20}\,
C^{ijkl}\,C_{ijkl}\right)
 \,,\label{CFTFW1}\\
\frac{\partial I_2}{\partial R^{i j}{}_{k l}}\,\tilde{\veps}^{i
j}\,\tilde{\veps}_{k l} &=&3\left(C^{k l m n}\, C_{m n}^{~~~ i j}\,
\tilde\veps_{i j}\,\tilde\veps_{k l} -C^{iklm}\,C^{j}_{~kl m
}\,\tilde{g}^{\perp}_{i j} + \frac{1}{5}\, C^{ijkl}\,C_{ijkl}\right)
 \,,\label{CFTFW2} \\
\frac{\partial I_3}{\partial R^{i j}{}_{k l}}\,\tilde{\veps}^{i
j}\,\tilde{\veps}_{k l} &=&  2\left(\Box\, C^{i j k l}+ 4\,R^{i}{}_{m}
C^{m j k l}-\frac{6}{5}\, R\, C^{i j k l}\right) \tilde\veps_{i
j}\,\tilde\veps_{k l}- 4\, C^{i j k l}\,R_{ik}\,\tilde{g}^{\perp}_{jl}
 \nonumber \\
&& \qquad+ 4\,C^{iklm}\,C^{j}_{~kl m }\,\tilde{g}^{\perp}_{i
j}-\frac{12}{5}\, C^{ijkl}\,C_{ijkl} \, .\label{CFTFW3}  
 \eea
Here $C_{ijkl}$ are the Weyl tensors, $l$ is the length scale of the entangling surface $\Sigma$ and $\delta$ is the short-distance cut-off that we use to regulate the calculations. $h_{ij}$ and $y^{a}$ are the induced metric and coordinates on the entangling surface $\Sigma$, respectively.  $\tilde{\veps}_{i j}$ and $\tilde{g}^{\perp}_{ij}$ are the two-dimensional volume form and metric in the space transverse to $\Sigma$, respectively. 

The logarithmic term of entanglement entropy can also be derived from the holographic entanglement entropy. We call this method as the holographic approach. Taking Einstein gravity as an example, the logarithmic term of entanglement entropy is given by \cite{Hung}
\be
S  =4\pi\log(\ell/\delta)\,
\int_{\Sigma}d^4y\sqrt{h}\left[\frac{1}{2}h^{ij} \gs_{ij} +
\frac{1}{8}(h^{ij} \go_{ij})^2 -\frac{1}{4}\go_{ij}\, h^{jk} \go_{kl}\,
h^{li}\right]\label{gogo}
\ee
where we have set the Newton's constant $G=\frac{1}{16\pi}$ the AdS radius $L=1$. The definitions of $g^{\scriptscriptstyle{(n)}}_{ij}$ can be found in the Fefferman-Graham expansion, i.e., $g_{ij}=g^{\scriptscriptstyle{(0)}}_{ij}+\rho g^{\scriptscriptstyle{(1)}}_{ij}+\rho^2g^{\scriptscriptstyle{(2)}}_{ij}+...\ $, for the asympotically Anti-de Sitter space 
\begin{eqnarray}\label{AdS}
ds^2=\frac{d\rho^2}{4\rho^2}+\frac{1}{\rho}g_{ij}(x,\rho)dx^idx^j.
\end{eqnarray}
Note that $x^i$ with $(i=1,2,...,6)$ are the coordinates on the boundary of AdS and $y^a$ with $(a=1,2,...,4)$ are the coordinates on the entangling surface $\Sigma$.

The mismatch between holographic result, eq.(\ref{gogo}), and field theoretical result, eq.(\ref{Waldformula8z}), becomes
 \bea\label{discrepancy}
\Delta S = -4\pi B_3\log(\ell/\delta)\, \int_{\Sigma}d^4y\sqrt{h}&(&C_{m n}{}^{r
s }C^{m n k l} \tilde{g}^\perp_{s l}\tilde{g}^\perp_{r k} -C_{m n
r}{}^s C^{m n r l}\tilde{g}^\perp_{s l}
   \nonumber \\
&&+ 2 C_{m}{}^n{}_r{}^s C^{mkrl} \tilde{g}^\perp_{n
s}\tilde{g}^\perp_{k l} - 2C_{m}{}^n{}_r{}^s C^{mkrl}
\tilde{g}^\perp_{n l}\tilde{g}^\perp_{k s})\,,
 \eea
This is the HMS mismatch \cite{Hung}. Note that the above equations are derived in the case of zero extrinsic curvatures. 

It is proposed to use the anomaly-like entropy and the generalized Wald entropy to explain the HMS mismatch in \cite{Miao1}. When the extrinsic curvatures vanish, only $C_{i jkl }^2C^{i jkl}\simeq -\nabla_m C_{ijkl}\nabla^m C^{ijkl}$ in $I_3$ contributes to non-zero anomaly-like entropy. Taking into account this contribution, the field theoretical and the holographic results match exactly. Note that the entropy of total derivative terms vanishes by applying the Lewkowycz-Maldacena regularization \cite{Maldacena,Dong1}.  However, the authors of \cite{Astaneh1,Astaneh2} claim that, in addition to $-\nabla_m C_{ijkl}\nabla^m C^{ijkl}$, the total derivative terms $B_3\nabla_m(C_{ijkl}\nabla^{m}C^{ijkl})+\nabla_i \hat{J}^i$ also contribute to the logarithmic term of entanglement entropy. They find that the entropy from total derivative terms is non-zero by applying the regularization of \cite{Fursaev}. And they suggest to utilize the entropy from total derivative terms to explain the HMS puzzle \cite{Astaneh1,Astaneh2}. Whether total derivative terms contribute to non-zero entropy or not is the main difference between \cite{Astaneh1,Astaneh2} and \cite{Miao1}. For simplicity, in this note we use the 'entropy from total derivative terms' to denote the contribution to the entanglement entropy which arises
from total derivative terms in the Weyl anomaly, i.e. from cohomologically trivial solutions to the Wess-Zumino consistency conditions.

\section{The proposal of \cite{Astaneh1,Astaneh2}}

In this section, we employ the method of \cite{Astaneh1,Astaneh2} to calculate the entropy from the total derivative terms in the holographic Weyl anomaly carefully. It turns out that the field theoretical result does not match the holographic one. Consequently, the proposal of \cite{Astaneh1,Astaneh2} does not solve the HMS puzzle. 

In the field theoretical approach, as we have explained in sect. 2, the logarithmic term of entanglement entropy can be derived from the Weyl anomaly  \cite{Hung, Fursaev}. In the case of Einstein gravity, the holographic Weyl anomaly is given by \cite{Henningson,Bastianelli}.
\begin{eqnarray}\label{WeylanomalyEinstein}
<T^i_{\ i}>=2\pi^3 E_6 -\frac{1}{16}I_1-\frac{1}{64}I_2+\frac{1}{192} (I_3-C_5)+\nabla_i J^i
\end{eqnarray}
where we have set the Newton's constant $G=\frac{1}{16\pi}$ and the AdS radius $L=1$. $C_5=\frac{1}{2}\Box C_{ijkl}C^{ijkl}$ and the total derivative term is given by \cite{Bastianelli}
\begin{eqnarray}\label{WeylanomalyTD1}
\nabla_i J^i=\frac{1}{960} (15 C_3-18 C_4-3 C_6+20 C_7)
\end{eqnarray}
with $C_k$ defined as
\begin{eqnarray}\label{WeylanomalyTD2ee}
C_3&=&\nabla_i[ R_{mn}\nabla^i R^{mn}-\frac{1}{6}R \nabla^i R]\\
C_4&=&\nabla_i[ R_{mn}\nabla^m R^{in}-\frac{1}{3}R^{im} \nabla^n R_{mn}-\frac{1}{18}R \nabla^i R]\\
C_6&=&\nabla_i[ \frac{1}{2}R^{im}\nabla_m R-R_{mn}\nabla^m R^{in}]\label{WeylanomalyTD2}\\
C_7&=&\nabla_i[ R^{kmni}\nabla_k R_{nm}+\frac{1}{4}R_{mnkl}\nabla^i R^{mnkl}+ \frac{1}{8}R_{im}\nabla_m R-\frac{1}{4}R_{mn}\nabla^m R^{in}].
\end{eqnarray}

In the case of zero extrinsic curvatures, the entropy of $E_6, I_1, I_2$ reduces to Wald entropy. Therefore, the HMS mismatch can only arise from $(I_3-C_5)$ and $\nabla_i J^i$.
Interestingly enough, although \cite{Miao1} and \cite{Astaneh2} take different choices of regularizations, they both agree that the total entropy minus the Wald entropy of $(I_3-C_5)$ can explain the HMS mismatch. To resolve the HMS puzzle completely, one needs to prove that the other total derivative terms  $\nabla_i J^i$ eq.(\ref{WeylanomalyTD1}) do not contribute to the entropy. Since the Lewkowycz-Maldacena regularization \cite{Maldacena} is used, the entropy of total derivative terms vanishes automatically in the approach of \cite{Miao1}.  This is, however, not the case for the approach of \cite{Astaneh1,Astaneh2}. As we shall show below, the entropy of the total derivative terms  eq.(\ref{WeylanomalyTD1}) is non-zero in their approach \cite{Astaneh2}. 

Now let us focus on the regularization \cite{Astaneh1,Astaneh2,Fursaev}. For simplicity, we apply the following regularized conical metric
\begin{eqnarray}\label{ConeAPS}
ds^2=f_{n}(r)dr^2+r^2d\tau^2+(\delta_{ab}+2\tilde{H}_{ab}r^{2n} \cos t \sin t) dy^ady^b,
\end{eqnarray}
where $f_n=\frac{r^2+b^2 n^2}{r^2+b^2}$ and $\tau\sim \tau+2n \pi$. Following the approaches of \cite{Astaneh1,Astaneh2}, we obtain
\begin{eqnarray}\label{APSTDss}
&&\int_{0}^{2\pi n}d\tau\int_{0}^{r_0}dr\int dy^{4}\sqrt{g}\nabla_i J^i=\int_{0}^{2\pi n}d\tau \int dy^{4}\sqrt{g}J^r|^{r= r_0}_{r=0}\\
&&=\int dy^{4}\frac{\pi}{40} (n-1) (bx)^{4(n-1)} \frac{x^8 (tr \tilde{H})^2+c_1 x^6+c_2x^4+c_3 x^2+c_4 }{(1+x^2)^4}|_{x=0}^{x=\infty}+O(n-1)^2\\
&&=\int dy^{4}\frac{\pi}{40} (n-1)(tr \tilde{H})^2+O(n-1)^2\label{APSTD}
\end{eqnarray}
where we have replaced $r$ by $bx$ in the above derivations. Note that \cite{Astaneh1,Astaneh2} choose to drop the contribution at $r=0\   (x=0) $. Thus $c_k$ are irrelevant to the final results. From eq.(\ref{APSTD}), we can derive the entropy of the total derivative term (\ref{WeylanomalyTD1}) as
\begin{eqnarray}\label{entropyTD}
S'_{\text{TD}}=-\lim_{n\to1}\partial_n(\int d\tau dr dy^4\sqrt{g}\nabla_i J^i)= -\frac{\pi}{40}\int dy^{4}(tr \tilde{H})^2
\end{eqnarray}
which is non-zero. Here `TD' means the total derivative terms. Note that eq.(\ref{entropyTD}) holds in the Lorentzian signature, which differs from its Euclidean form by a minus sign. Now it is clear that the proposal of \cite{Astaneh1,Astaneh2} can not solve the HMS puzzle \cite{Hung}. In other words, the field theoretical and the holographic results of the logarithmic term of entanglement entropy fail to match in the approach of \cite{Astaneh1,Astaneh2}. However, it is not surprising. As we know, the total derivative terms in the Weyl anomaly come from cohomologically trivial solutions to the Wess-Zumino consistency conditions, accodingly, there is no reason that they could contribute to the entropy.

\section{The proposal of \cite{Miao1}}

In this section, we prove that the entropy of the total derivative eq.(\ref{WeylanomalyTD1}) indeed vanishes in the approach of \cite{Miao1}, therefore, the proposal of \cite{Miao1} solves the HMS puzzle \cite{Hung}. We apply Lewkowycz-Maldacena regularization \cite{Maldacena,Dong1} instead of the regularization \cite{Fursaev} in this section. 

Let us focus on the following regularized conical metric
\begin{eqnarray}\label{ConeDong}
ds^2=\frac{1}{(r^2+b^2)^{1-\frac{1}{n}}}(dr^2+r^2d\tau^2)+(\delta_{ab}+2\tilde{H}_{ab}r^{2} \cos t \sin t)dy^a dy^b
\end{eqnarray}
with $\tau\sim \tau+2\pi$. For the total derivative eq.(\ref{WeylanomalyTD1}),  we firstly expand it in powers of $\tilde{H}$ and then perform the $\tau$ integral. It turns out that only the $\tilde{H}^2$ terms contribute to the entropy. The other terms are either in higher order $O(n-1)^2$ or vanishing in the limit $b\to 0$. Focus on the $\tilde{H}^2$ terms, we have
\begin{eqnarray}\label{TDDong}
&&\int_{0}^{2\pi}d\tau\int_{0}^{r_0}dr\int dy^4\sqrt{g} \nabla_i J^i\nonumber\\
&=&b^{4-\frac{4}{n}} \int_0^{\infty}dx\int dy^4\frac{\pi  x}{60 \left(x^2+1\right)^{\frac{2}{n}+4}}\big{ [} (n-1)\sum_{k=0}^{4} q_{2k} x^{2k} \nonumber\\
&&+ (n-1)^2[ (40tr \tilde{H}^2-10(tr \tilde{H})^2)x^{10}+\sum_{k=0}^4 p_{2k} x^{2k} ]+ O(n-1)^3\big{]}
\end{eqnarray}
where $r= bx$ and $q_{2k}$ are given by
\begin{eqnarray}\label{QQDong}
&&q_0=28tr \tilde{H}^2-(tr \tilde{H})^2,\ \ q_2=12(10tr \tilde{H}^2-3(tr \tilde{H})^2),\\
&&q_4=2(78tr \tilde{H}^2-33(tr \tilde{H})^2),\ \ q_6=4(16tr \tilde{H}^2-7(tr \tilde{H})^2),\\
&&q_{8}=3(tr\tilde{H})^2
\end{eqnarray}
$p_{2k}$ are irrelevant to the entropy. We find the following formulae are useful \footnote{In principle, one should firstly intrgrate x from 0 to $(r_0/b)$ and then subtract off the contributions from the singular cone with $b=0$. The detailed approach can be found in \cite{Dong2}. In general, there are non-universal terms which depend on $r_0$ and the universal terms in the integral. Only the universal terms survive once we subtract off the contributions from the singular cone. Here we use a simpler method. We integrate x from 0 to $\infty$ for some suitable range of n, and then do the analytic continuation for n. It turns out that only the universal terms appear in the results. Thus the method here produces the same results as the one of \cite{Dong2}.}
\begin{eqnarray}\label{TDWaldDong}
&&\int_{0}^{\infty}\frac{xdx}{\left(1+x^2\right)^{4+\frac{2}{n}}}=\int_{0}^{\infty}\frac{x^9dx}{\left(1+x^2\right)^{4+\frac{2}{n}}}=\frac{1}{10} +O(n-1)\\
&&\int_{0}^{\infty}\frac{x^3dx}{\left(1+x^2\right)^{4+\frac{2}{n}}}=\int_{0}^{\infty}\frac{x^7dx}{\left(1+x^2\right)^{4+\frac{2}{n}}}=\frac{1}{40} +O(n-1)\\
&&\int_{0}^{\infty}\frac{x^5dx}{\left(1+x^2\right)^{4+\frac{2}{n}}}=\frac{1}{60} +O(n-1)\\
&&\int_{0}^{\infty}\frac{x^{11}dx}{\left(1+x^2\right)^{4+\frac{2}{n}}}=\frac{60 \Gamma \left(\frac{2}{n}-2\right)}{\Gamma \left(4+\frac{2}{n}\right)}=-\frac{1}{4(n-1)}+O(n-1)^0
\end{eqnarray}
From the above equations together with $b^{4-\frac{4}{n}} =1+O(n-1)$, we can derive the entropy of the total derivative term ( \ref{WeylanomalyTD1}) as
\begin{eqnarray}\label{TDDong1}
S''_{\text{TD}}=\lim_{n\to1}\partial_n(\int d\tau dr dy^4\sqrt{g}\nabla_i J^i)=0,
\end{eqnarray}
which indeed vanishes. Here `TD' denotes the total derivative.  Now it is clear that the proposal of \cite{Miao1} indeed resolve the HMS puzzle \cite{Hung}.

\section{Double checks}

In this section, we provide further support that it is the proposal of \cite{Miao1} rather than the one of \cite{Astaneh1,Astaneh2} that can resolve the HMS puzzle. We calculate the entropy for all the terms in the Weyl anomaly eq.(\ref{WeylanomalyEinstein}) by using the methods of \cite{Miao1} and \cite{Astaneh1,Astaneh2}, respectively. It turns out that only the method of \cite{Miao1} can yield consistent result with the holographic one. This can be regarded as a double check of the results of sect. 3 and sect. 4. 

In the holographic approach, the universal term of entanglement entropy for 6d CFTs dual to Einstein gravity is given by \cite{Hung}
\begin{eqnarray}\label{HoloLog}
S  =4\pi\log(\ell/\delta)\,
\int_{\Sigma}d^4y\sqrt{h}\left[\frac{1}{2}h^{ij} \gs_{ij} +
\frac{1}{8}(h^{ij} \go_{ij})^2 -\frac{1}{4}\go_{ij}\, h^{jk} \go_{kl}\,
h^{li}\right]
\end{eqnarray}
The above formula applies to the case with zero extrinsic curvatures. For the general case, please see \cite{Miao2}. For the conical metrics eqs.(\ref{ConeAPS},\ref{ConeDong}) with $b=0$ and $n=1$, the above equation becomes 
\begin{eqnarray}\label{HoloLogCone}
S=-\frac{\pi}{40}\log(\ell/\delta)\int_{\Sigma} d^4y \sqrt{h}[ 8 tr \tilde{H}^2-(tr \tilde{H} )^2 ]
\end{eqnarray}

Let us rewrite the holographic Weyl anomaly eq.(\ref{WeylanomalyEinstein1}) in the initial form of \cite{Henningson}
\begin{eqnarray}\label{WeylanomalyEinstein1}
<T^i_{\ i}>=\frac{1}{32}\big{(} -\frac{1}{2}RR^{ij}R_{ij}+\frac{3}{50}R^3+R^{ij}R^{kl}R_{ikjl}-\frac{1}{5}R^{ij}\nabla_i\nabla_j R+\frac{1}{2}R^{ij}\Box R_{ij}-\frac{1}{20}R\Box R  \big{)}
\end{eqnarray}
Note that the curvature in our notation is different from the one of \cite{Henningson} by a minus sign. 

Using the method of \cite{Astaneh1,Astaneh2,Fursaev} together with the metric eq.(\ref{ConeAPS}), we derive the total entropy of eq.(\ref{WeylanomalyEinstein1}) in the Lorentzian signature as
\begin{eqnarray}\label{APSLog}
S'=-\frac{\pi}{5}\int_{\Sigma} d^4y \sqrt{h}[ tr \tilde{H}^2 ],
\end{eqnarray}
which does not match the holographic result eq.(\ref{HoloLogCone}) at all. While applying the approach of \cite{Miao1, Dong1} with the conical metric eq.(\ref{ConeDong}), we obtain the total entropy of eq.(\ref{WeylanomalyEinstein1}) as
\begin{eqnarray}\label{MGLog}
S''=-\frac{\pi}{40}\int_{\Sigma} d^4y \sqrt{h}[ 8 tr \tilde{H}^2-(tr \tilde{H} )^2 ]
\end{eqnarray}
which exactly agrees with the holographic result eq.(\ref{HoloLogCone}). Please refer to the Appendix for the derivations of eqs.(\ref{APSLog},\ref{MGLog}). Recall that the entropies of $E_6, I_1, I_2$ and $(I_3-C_5)$ are the same in the approaches of \cite{Miao1} and \cite{Astaneh1,Astaneh2} when the extrinsic curvatures vanish.  And the only difference of the entropy in these two approaches comes from the total derivative term $\nabla_i J^i$. Thus it is expected that we have $S''_{\text{TD}}-S'_{\text{TD}}=S''-S'$. From eqs.(\ref{entropyTD},\ref{TDDong1},\ref{APSLog},\ref{MGLog}), we find that this is indeed the case. This can be regarded as a check of our calculations. Now it is clear that it is the proposal of \cite{Miao1} rather than the one of \cite{Astaneh1,Astaneh2} that can resolve the HMS puzzle. 

\section{The arbitrariness of Wald entropy formula}

In this section, we show that the Wald entropy formula is not well-defined. In general, it is inconsistent with the Bianchi identities. However, this is not surprising. In addition to Wald entropy, the anomaly-like entropy \cite{Fursaev,Dong1,Camps} and the generalized Wald entropy \cite{Miao1} also contribute to the total entropy. It does not matter as long as the total entropy is well-defined. As we shall show below, this is indeed the case. Note that the arbitrariness of Wald entropy does not affect our discussions above, since we always focus on the total entropy in this note. It is found that there is arbitrariness in the Noether charge method on horizon with non-zero extrinsic curvatures \cite{Jacobson,Wald2}. And the Wald entropy formula
\begin{eqnarray}\label{Waldentropy}
S_{\text{Wald}}=-2\pi \int_{\Sigma}dy^{D-2}\sqrt{h}\frac{\delta L}{\delta R_{ijkl}} \tilde{\veps}_{ij}\tilde{\veps}_{kl}
\end{eqnarray}
is just one of the several possible candidates for the entropy. The observation of this section is a little different. We notice that, in general, the Wald entropy formula eq.(\ref{Waldentropy}) gives different results for equivalent actions. It is necessary to point out this subtlety.

Let us take an example to illustrate the arbitrariness of the Wald entropy formula eq.(\ref{Waldentropy}).
We work in Euclidean signature in this section. Thus we have $\tilde{\veps}_{ij}\tilde{\veps}^{ij}=2,\   \tilde{\veps}^{im}\tilde{\veps}^{j}_{\ m}=\tilde{g}^\perp{}^{i j}$.
From the Bianchi identities, we have
\begin{eqnarray}\label{example1}
\frac{1}{4}\nabla_i R\nabla^i R=\nabla_i R^{im}\nabla_j R^{j}_{\ m}
\end{eqnarray}
The Wald entropy of the left hand side of eq.(\ref{example1}) is given by
\begin{eqnarray}\label{examplewald11}
2\pi\int_{\Sigma}dy^{D-2}\sqrt{h}\Box R 
\end{eqnarray}
While the Wald entropy of the right hand side of eq.(\ref{example1}) can be derived as
\begin{eqnarray}\label{examplewald12}
2\pi\int_{\Sigma}dy^{D-2} \sqrt{h}\tilde{g}^\perp{}^{i j}\nabla_i\nabla_j R =2\pi\int_{\Sigma}dy^{D-2}\sqrt{h} [ \Box R -D_i D^i R+k^a\nabla_a R ]
\end{eqnarray}
where $D_i$ are the intrinsic covariant derivatives and $k^a=k^{a}_{\ ij}g^{ij}$ are the traces of the extrinsic curvatures. Clearly, eq.(\ref{examplewald11}) and eq.(\ref{examplewald12}) are different for the case with non-zero extrinsic curvatures. This implies that, in general, the Wald entropy is not a well-defined physical quantity. It should be mentioned that Wald entropy works well for entangling surfaces $\Sigma$ with the rotational symmetry. Thus nothing goes wrong in the initial work of Wald \cite{Wald}. For entangling surfaces $\Sigma$ with the rotational symmetry,  Wald entropy becomes the total entropy and thus must be well-defined. 

The total entropy of left hand side and the right hand side of eq.(\ref{example1}) can be calculated by using the method of the appendix. Clearly, both sides give the same results. That is because eq.(\ref{example1}) is an identity, therefore the left hand side and the right hand side of eq.(\ref{example1}) make no differences in the approach of the appendix. This implies only the total entropy is well-defined. On the contrary, there is arbitrariness in the derivations of the Wald entropy. The Wald entropy changes when one rewrite the action into an equivalent form by using the Bianchi identities.  

Let us consider another example. Let us rewrite the total derivative $C_6$ eq.(\ref{WeylanomalyTD2}) into two equivalent expressions. The first one is 
\begin{eqnarray}\label{C61}
\bar{C}_6=\frac{1}{4}\nabla_i R\nabla^i R-\nabla_i R_{mn}\nabla^mR^{in}+ R^{ij}\nabla_i\nabla_j R-2R^{ij} \nabla_{(i}\nabla_{k)}R^{k}_{\ j} 
\end{eqnarray}
and the second one is \cite{Astaneh3}
\begin{eqnarray}\label{C62}
\hat{C}_6=\frac{1}{4}\nabla_i R\nabla^i R-\nabla_i R_{mn}\nabla^mR^{in}+R_{ij}R_{kl}R^{ikjl}-R^i_{\ j}R^j_{\ k}R^k_{\ i}
\end{eqnarray} 
After some calculations, we derive the Wald entropy of $\bar{C}_6$ and $\hat{C}_6$ as
\begin{eqnarray}\label{WaldC61}
&&\bar{S}_{\text{Wald}}=0,\\
&&\hat{S}_{\text{Wald}}=2\pi\int_{\Sigma} dy^{D-2}\sqrt{h}[ \Box R-\tilde{g}^\perp{}^{i j}\nabla_i\nabla_j R +\tilde{g}^\perp{}^{i j}R_{im}R^m_{\ j}-R_{ij}R_{kl}(\tilde{g}^\perp{}^{i j}\tilde{g}^\perp{}^{k l}-\tilde{g}^\perp{}^{i l}\tilde{g}^\perp{}^{k j}) ] \label{WaldC62}\nonumber\\
\end{eqnarray}
Remarkably, although the total derivatives $\bar{C}_6$ and $\hat{C}_6$ are equivalent, they give different Wald entropy \footnote{Eq.(\ref{WaldC62}) is derived independently in a recent work \cite{Astaneh3}. However, they do not realize that there is arbitrariness in the derivations of Wald entropy.}. This clearly shows that Wald entropy is not a well-defined physical quantity. There is too much arbitrariness in its derivations. On the other hand, the total entropy is indeed well defined. One can check that the total entropy of $\hat{C}_6$ and $\bar{C}_6$ is both zero by using the approach of \cite{Miao1}. By applying the approach of \cite{Astaneh1,Astaneh2} instead, the total entropy of $\hat{C}_6$ and $\bar{C}_6$ is non-zero, but still the same. 

In conclusion, the Wald entropy makes no sence by itself. There is too much arbitrariness in its derivations. Instead, only the total entropy consisted of Wald entropy\cite{Wald}, the generalized Wald entropy \cite{Miao1} and the anomaly-like entropy \cite{Fursaev,Dong1,Camps} is well-defined.

\section{Conclusion}

In this note, we have discussed two possible resolutions to the HMS entropy discrepancy \cite{Hung}. By studying the CFT dual to Einstein gravity, we find that it is the proposal of \cite{Miao1} rather than the one of \cite{Astaneh1,Astaneh2} that can resolve the HMS puzzle. This implies that the Lewkowycz-Maldacena regularization \cite{Maldacena,Dong1} instead of the regularization \cite{Astaneh1,Astaneh2,Fursaev} yields the correct results for the entropy. It is a strong support for the work \cite{Dong2} that the covariant total derivative terms do not contribute to non-zero entropy. Finally, we show that the Wald entropy formula is not well-defined, since in general it gives different results for equivalent actions. It turns out that only the total entropy is well-defined. Notice that in stationary space-times Wald entropy becomes the total entropy and thus is well-defined, which suggests nothing goes wrong in the initial work of Wald \cite{Wald}.

\section*{Acknowledgements}

R. X. Miao is supported by Sino-Germann (CSC-DAAD) Postdoc Scholarship Program. Y. Huang and R. X. Miao thank School of Astronomy and Space Science at Sun Yat-Sen University for hospitality. 

\appendix

\section{Detailed calculations}

In this appendix, we derive eqs.(\ref{APSLog},\ref{MGLog}) with some details.

By using the FPS regularization eq.(\ref{ConeAPS}), we can derive eq. (\ref{APSLog}). Firstly, we expand the holographic Weyl anomaly eq.(\ref{WeylanomalyEinstein1}) in powers of $\tilde{H}$ and then do the $\tau$ integral. Here we take $n$ as an integer.  Secondly, we do the analytic continuation for $n$ and expand the results around $n=1$. We keep terms up to the order $O(n-1)^2$. Finally, we do the r integral and select the terms in order $O(n-1)$. It turns out that only the $\tilde{H}^2$ terms contribute to the entropy. The other terms are either in higher order $O(n-1)^2$ or vanishing in the limit $b\to 0$. Focus on the $\tilde{H}^2$ terms, we have
\begin{eqnarray}\label{APSLogcalculation}
\int drd\tau dy^4 \sqrt{g} <T^i_{\ i}>&=&b^{4 n-4} \int_0^{\infty}dx\int dy^4\frac{\pi  \left(x^{4 n-3}\right)}{20 \left(x^2+1\right)^6} \big{ [} (n-1)  \sum_{k=0}^{4} d_{2k} x^{2k} \nonumber\\
&&+ (n-1)^2 [ \sum_{k=0}^{4} c_{2k} x^{2k} + 2(3(tr\tilde{H})^2-10tr \tilde{H}^2) x^{10} ]\nonumber\\
&& + O(n-1)^3 \big{]}
\end{eqnarray}
where $r= bx$ and $d_{2k}$ are given by
\begin{eqnarray}\label{APSdk1}
&&d_0=4tr \tilde{H}^2, \ \ d_2=6(tr\tilde{H})^2+16tr \tilde{H}^2,\\
&&d_4=9 ((tr\tilde{H})^2+4 tr \tilde{H}^2),\ d_6=40tr \tilde{H}^2,\\
&&d_8=-3(tr\tilde{H})^2+16 tr \tilde{H}^2
\end{eqnarray}
Note that $c_{2k}$ are irrelevant to the final result, so we do not list them. The first line of eq.(\ref{APSLogcalculation}) contribute to the Wald-like entropy. The second line of eq.(\ref{APSLogcalculation}) are the would-be logarithmic terms. Naively, second line of eq.(\ref{APSLogcalculation}) is in order $O(n-1)^2$. It seems to be irrelevant to the entropy. However, it becomes in order $O(n-1)$ after the integral.  The magic happens because the would-be logarithmic divergence gets a $\frac{1}{n-1}$ enhancement.

In general, we have two kind of would-be logarithmic terms. One is at $x\to0$ and the other one is at $x\to\infty$. 
\begin{eqnarray}\label{wouldlogAPS1}
\int_{0}^{\infty}\frac{x^{4 n-5}dx}{\left(1+x^2\right)^6}&=&-\frac{1}{60} \pi  (n-3) (n-2) (2 n-7) (2 n-5) (2 n-3) \csc (2 \pi  n), \ \ 1<\Re(n)<4\nonumber\\
&=&\frac{1}{4 (n-1)}+O(n-1)^0\\
\int_{0}^{\infty}\frac{x^{4 n+7}dx}{\left(1+x^2\right)^6}&=&-\frac{1}{60} \pi  n (n+1) (2 n-1) (2 n+1) (2 n+3) \csc (2 \pi  n),\ \ -2<\Re(n)<1\label{wouldlog1}\nonumber\\
&=&\frac{-1}{4(n-1)}+O(n-1)^0,\label{wouldlogAPS2}
\end{eqnarray}
It seems that the above two integrals could not be well-defined at the same time. Thus, the authors of \cite{Astaneh1,Astaneh2} choose to drop the would-be logarithmic term at infinity eq.(\ref{wouldlogAPS2}). However, as pointed out in \cite{Dong2}, we actually do not need the condition $n<1$ to derive eq.(\ref{wouldlogAPS2}). Note also that the results after analytic continuation are both well defined for $n<1$ and $n>1$. So there is no reason to drop such term. However, since we are using the methods of \cite{Astaneh1,Astaneh2},  we adopt their choice in this paper. Note that the would-be logarithmic term at $x\to0$ vanishes in our case.   In addition to eq.(\ref{wouldlogAPS1}), we find the following formulas are useful
\begin{eqnarray}\label{WaldAPS00}
&&\int_{0}^{\infty}\frac{x^{4 n-3}dx}{\left(1+x^2\right)^6}=\int_{0}^{\infty}\frac{x^{4 n+5}dx}{\left(1+x^2\right)^6}=\frac{1}{10} +O(n-1)\\
&&\int_{0}^{\infty}\frac{x^{4 n-1}dx}{\left(1+x^2\right)^6}=\int_{0}^{\infty}\frac{x^{4 n+3}dx}{\left(1+x^2\right)^6}=\frac{1}{40} +O(n-1)\\
&&\int_{0}^{\infty}\frac{x^{4 n+1}dx}{\left(1+x^2\right)^6}=\frac{1}{60}+O(n-1)\label{WaldAPS}
\end{eqnarray}
Using eqs.(\ref{wouldlogAPS1}-\ref{WaldAPS}) together with $b^{4n-4}=1+O(n-1)$, we can derive 
\begin{eqnarray}\label{APSLogcalculation1}
\int drd\tau dy^4 \sqrt{g} <T^i_{\ i}>&=&\frac{(n-1)\pi }{5}\int dy^4 [ tr \tilde{H}^2 ]+ O(n-1)^2
\end{eqnarray}
Now we get the entropy eq.(\ref{APSLog}) in Lorentzian signature 
\begin{eqnarray}\label{APSLogA}
S_{\text{APS}}=-\lim_{n\to1}\partial_n \int drd\tau dy^4 \sqrt{g} <T^i_{\ i}>=-\frac{\pi}{5}\int_{\Sigma} d^4y \sqrt{h_0}[ tr \tilde{H}^2 ].
\end{eqnarray}
Note that the entropy in Lorentzian signature differs from its Euclidean form by a minus sign. In the above derivations we have dropped the would-be logarithmic term at $x\to\infty$ as \cite{Astaneh1,Astaneh2}. Even if we recover this kind of term, the field theoretical result still does not match the holographic one. 

Now let us turn to derivation of eq.(\ref{MGLog}). The calculation is very similar to the above one. The only difference is that now we use Dong's regularization for the conical metric eq.(\ref{ConeDong}).  We obtain 
\begin{eqnarray}\label{MGLogcalculation}
\int drd\tau dy^4 \sqrt{g} <T^i_{\ i}>=b^{4-\frac{4}{n}} \int_0^{\infty}dx\int dy^4\frac{\pi  x}{100 \left(x^2+1\right)^{\frac{2}{n}+4}}\big{ [} (n-1)\sum_{k=0}^{4} f_{2k} x^{2k} + O(n-1)^2 \big{]}
\end{eqnarray}
where $r= bx$ and $f_{2k}$ are given by
\begin{eqnarray}\label{APSdk}
&&f_0=5 (3(tr\tilde{H})^2-4tr \tilde{H}^2),\ \ f_2=-80 tr \tilde{H}^2,\\
&&f_4=-30((tr\tilde{H})^2+6tr \tilde{H}^2),\ \ f_6=-200 tr \tilde{H}^2\\
&&f_{8}=15(tr\tilde{H})^2-80tr \tilde{H}^2
\end{eqnarray}
The following formulas are useful
\begin{eqnarray}\label{WaldDong00}
&&\int_{0}^{\infty}\frac{xdx}{\left(1+x^2\right)^{4+\frac{2}{n}}}=\int_{0}^{\infty}\frac{x^9dx}{\left(1+x^2\right)^{4+\frac{2}{n}}}=\frac{1}{10} +O(n-1)\\
&&\int_{0}^{\infty}\frac{x^3dx}{\left(1+x^2\right)^{4+\frac{2}{n}}}=\int_{0}^{\infty}\frac{x^7dx}{\left(1+x^2\right)^{4+\frac{2}{n}}}=\frac{1}{40} +O(n-1)\\
&&\int_{0}^{\infty}\frac{x^5dx}{\left(1+x^2\right)^{4+\frac{2}{n}}}=\frac{1}{60} +O(n-1)\label{WaldDong}
\end{eqnarray}
From eqs.(\ref{MGLogcalculation}-\ref{WaldDong}), we can derive 
\begin{eqnarray}\label{MGLogcalculation1}
\int drd\tau dy^4 \sqrt{g} <T^i_{\ i}>=-\frac{(n-1)\pi}{40}\int d^4y [ 8 tr \tilde{H}^2-(tr \tilde{H} )^2 ]+O(n-1)^2
\end{eqnarray}
Now we obtain the entropy eq.(\ref{MGLog}) in the Lorentzian signature
\begin{eqnarray}\label{MGLogA}
S_{\text{MG}}=\lim_{n\to1}\partial_n \int drd\tau dy^4 \sqrt{g} <T^i_{\ i}>=-\frac{\pi}{40}\int_{\Sigma} d^4y \sqrt{h_0}[ 8 tr \tilde{H}^2-(tr \tilde{H} )^2 ].
\end{eqnarray}
Note that Dong's formula of entropy (the first equality of eq.(\ref{MGLogA})) \cite{Dong1} differs from the one of FPS  (the first equality of eq.(\ref{APSLogA})) \cite{Fursaev} by a minus sign.

\end{document}